\documentclass[aps,pra,floatfix,twocolumn,superscriptaddress]{revtex4}

\usepackage{graphicx}
\usepackage{graphics}
\usepackage{amssymb}
\usepackage{amsmath} 
\usepackage{epsfig}
\usepackage{latexsym}
\usepackage{color}
\usepackage{subfigure}
\usepackage{array}
\usepackage{bm}

\newcommand{\ket}[1]{|#1\rangle}

\newcommand{\braket}[2]{\langle #1|#2\rangle}
\newcommand{\ketbra}[2]{|#1\rangle\langle#2|}
\newcommand{\braopket}[3]{\langle #1|#2|#3\rangle}

\begin{document}

\title{Comment on `Grover Search with Pairs of Trapped Ions'}

\author{Charles D. Hill} 
\email{hillcd@physics.uq.edu.au}
\affiliation{Centre for Quantum Computer
Technology, and Department of Physics, The University of Queensland, St
Lucia, QLD 4072, Australia}

\author{Hsi-Sheng Goan}
\email{goan@physics.uq.edu.au}
\affiliation{Centre for Quantum Computer Technology, University of
New South Wales, Sydney, NSW 2052, Australia}
\thanks{Mailing Address: Centre for Quantum Computer
Technology, C/- Department of Physics, The University of Queensland, St
Lucia, QLD 4072, Australia}


\begin{abstract}
In this brief comment on `Grover search with pairs of trapped Ions'
[Phys. Rev. A \textbf{63}, 052308, (2001)], we show that Grover's
algorithm may be performed exactly using the gate set given provided
that small changes are made to the gate sequence. An analytic
expression for the probability of success of Grover's algorithm for
any unitary operator, U, instead of Hadamard is presented.
\end{abstract}

\maketitle


In the paper `Grover search with trapped ions' \cite{Fen01}, M. Feng
attempts to describe how Grover's algorithm may be performed using
trapped ions. In contrast to earlier proposals, Feng proposes using
pairs of trapped ions. The advantage of this scheme is that it would
eliminate one source of phase error known as superpositional wave
function oscillations \cite{BDT00}. However, we think the incorrect
gate sequence for Grover's algorithm is used in Ref. \cite{Fen01}, and
therefore the results obtained are not ideal. By making a small change
to the gate sequence given, we show that Grover's algorithm may be
performed exactly using only the operations introduced by Feng.

Two quantum operations are introduced by Feng \cite{Fen01}. The first
operation is an X rotation, notated by $U$,
\begin{eqnarray}
U(\theta) &=&  R_x\left(2\theta\right)\\
  	  &=& \left( 
		\begin{array}{cc}
		\cos{\theta}   & -i \sin{\theta} \\
		-i \sin{\theta} & \cos{\theta} \\
		\end{array}
	      \right).
\end{eqnarray}
A derivation for this operation may be found in Ref. \cite{SM99}. To
create a rotation by $\frac{\pi}{2}$ the authors suggest rotating by
$\theta = 7\pi/4$, although this rotation may be obtained more easily
by rotating by $\theta = 3\pi/4$. For clarity we define
\begin{equation}
W_n = R_x\left(-\frac{\pi}{2}\right)^{\otimes n}. \label{eqn:W}
\end{equation}
as in Eq. (6) of Ref. \cite{Fen01}, where $n$ is the number of qubits.

The second gate operation introduced is the $\Lambda_1Y$ (controlled
Y) operation. This operation is denoted by $M$ in Feng's original
paper:
\begin{eqnarray}
M_1^{(2)} = \left( 
	\begin{array}{cccc}
	1 & 0 & 0 & 0 \\
	0 & 1 & 0 & 0 \\
	0 & 0 & 0 & -i \\
	0 & 0 & i & 0
	\end{array}
	\right).
\end{eqnarray}

This operation may be used to create entanglement. It is not explicitly
stated how this operation or the multi-qubit $\Lambda_nY$ operation is
performed. However, using this operation, it is clear that it is
possible to change the sign of one state, as is desirable in Grover's
algorithm. This is described by the operation $P_m^{(n)}$ where $n$ is
the number of qubits, and $m$ is the number of the state to change
sign. Using this gate set ($U(\theta)$ and $M$) Feng attempts to
perform Grover's algorithm, which as he correctly realizes should be
possible.

The major issue in Feng's \cite{Fen01} paper, we think, is to
incorrectly implement Grover's algorithm. This is masked by the fact
that the graphs given in Feng's paper (Fig. 1, Fig. 2, and Fig. 3) are
labeled incorrectly, and show amplitude, not probability. The maximum
probability of success obtained in the search, using the method given
in \cite{Fen01}, for the $\ket{111}$ state was approximately $38\%$,
and not the $62\%$ shown on the graph. The actual probabilities for
his implementation can be found by squaring amplitudes given, making
all of the probabilities of success, considerably less than those
quoted.

A (correct) prescription for performing such a search is given by
Grover\cite{Gro98} which may be implemented using $X$ rotations
instead of Hadamard gates. To correctly perform Grover's algorithm
requires rotations of both $R_x(\frac{\pi}{2})$ and
$R_x(-\frac{\pi}{2})$. These are not implemented in Feng's
paper. Specifically, Eq. (4) in Ref. \cite{Fen01} should be specified
by
\begin{equation}
D_2 = W_2 P^{(2)}_1 W_2^{\dagger},
\end{equation}
and not
\begin{equation}
D_2 = W_2 P^{(2)}_1 W_2.
\end{equation}

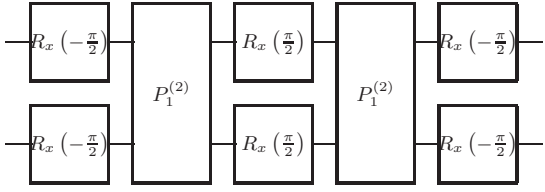
\begin{figure}[ht!]
\begin{center}
\scalebox{0.85}{


\unitlength 0.42mm

\begin{picture}(210, 86)

\linethickness{0.1mm}
\put(0, 24) {\line(1,0) {10}}
\put(0, 62) {\line(1,0) {10}}
\linethickness{0.4mm}
\put(10, 10){\framebox(28, 28){$R_{x} \left(-\frac{\pi}{2}\right)$}}
\linethickness{0.4mm}
\put(10, 48){\framebox(28, 28){$R_{x} \left(-\frac{\pi}{2}\right)$}}
\linethickness{0.1mm}
\linethickness{0.1mm}
\put(38, 24) {\line(1,0) {10}}
\put(38, 62) {\line(1,0) {10}}
\linethickness{0.4mm}
\put(48, 10){\framebox(28, 66){$P_{1}^{(2)}$}}
\linethickness{0.1mm}
\linethickness{0.1mm}
\put(76, 24) {\line(1,0) {10}}
\put(76, 62) {\line(1,0) {10}}
\linethickness{0.4mm}
\put(86, 10){\framebox(28, 28) {$R_{x} \left(\frac{\pi}{2}\right)$}}
\linethickness{0.4mm}
\put(86, 48){\framebox(28, 28) {$R_{x} \left(\frac{\pi}{2}\right)$}}
\linethickness{0.1mm}
\linethickness{0.1mm}
\put(114, 24) {\line(1,0) {10}}
\put(114, 62) {\line(1,0) {10}}
\linethickness{0.4mm}
\put(124, 10){\framebox(28, 66){$P_{1}^{(2)}$}}
\linethickness{0.1mm}
\linethickness{0.1mm}
\put(152, 24) {\line(1,0) {10}}
\put(152, 62) {\line(1,0) {10}}
\linethickness{0.4mm}
\put(162, 10){\framebox(28, 28){$R_{x} \left(-\frac{\pi}{2}\right)$}}
\linethickness{0.4mm}
\put(162, 48){\framebox(28, 28){$R_{x} \left(-\frac{\pi}{2}\right)$}}
\linethickness{0.1mm}
\linethickness{0.1mm}
\put(190, 24) {\line(1,0) {10}}
\put(190, 62) {\line(1,0) {10}}

\end{picture}

}
\caption{Gate Sequence of Grover Search on Two Qubits} \label{fig:groverCirc}
\end{center}
\end{figure}

Figure \ref{fig:groverCirc} shows the corrected circuit diagram for
Grover's algorithm acting on 2 qubits. For larger systems of more than
two qubits, similar circuits can be drawn.

Although our change makes no difference to probability of success of
two qubit states, the corresponding change for states of three or more
qubits makes a significant difference to the success of the
algorithm. The probability of finding a marked state may be found
analytically for the modified gate sequences, as shown in the later in
this paper. The probabilities exhibit the same periodic behavior, and
have the same maximum probabilities expected from Grover's
algorithm. With this small change to the gate sequence, Grover's
algorithm may be performed exactly using the operations introduced by
Feng.



In fact, a simple analytic expression can be derived for the
probability of success of Grover's algorithm using an arbitrary
rotation, $U$, in the place of the Hadamards. Consider one step of the
algorithm, given by
\begin{equation}
Q = -P_{\gamma} U^{\dagger} P_{\tau} U. \label{eqn:groverIt}
\end{equation}
In this equation, $\ket{\gamma}$ is the initially prepared state whose
sign is also flipped by the Grover iteration. $\ket{\tau}$ is the
marked state. The subspace spanned by $\ket{\gamma}$ and $U^{\dagger}
\ket{\tau}$ is invariant under the operation $Q$. Using the identities
\begin{eqnarray}
P_{\gamma} &=& I - 2 \ketbra{\gamma}{\gamma}, \\
P_{\tau}   &=& I - 2 \ketbra{\tau}{\tau},
\end{eqnarray}
it can be shown \cite{Gro98} that
\begin{eqnarray}
Q\left[ \begin{array}{c}
	\ket{\gamma} \\
	U^{\dagger} \ket{\tau}
	 \end{array}\right] =
	\left[ \begin{array}{cc}
	(1-4|U_{\tau \gamma}|^{2}) & 2 U_{\tau \gamma} \\
	-2 U_{\tau \gamma}^{*} & 1
	 \end{array}\right]
\left[ \begin{array}{c}	
	\ket{\gamma} \\
	U^{\dagger} \ket{\tau}
	\end{array}
\right],
\end{eqnarray}
where
\begin{equation}
U_{\tau\gamma} = \braopket{\tau}{U}{\gamma}.
\end{equation}
Notice that $U^{\dagger} \ket{\tau}$ and $\ket{\gamma}$ are not
orthogonal. We wish to represent the rotation $Q$ in an orthonormal
basis. To do this we introduce the vector
\begin{equation}
\ket{\gamma'} = \frac{\ket{\gamma} - U_{\tau\gamma}
U^{\dagger}\ket{\tau}}{\sqrt{1-|U_{\tau\gamma}|^{2}}}. \label{eqn:gamma}
\end{equation}
$\ket{\gamma'}$ and $U^{\dagger}\ket{\tau}$ form an effective spin
orthonormal basis for the subspace on which $Q$ acts. That is,
$\braket{\gamma'}{\gamma'} = 1$ and $\braopket{\gamma'}{U^{\dagger}}
{\tau} = 0$.

We now show that in this new basis, the Grover iteration simply
represents a rotation in $SU(2)$. An arbitrary rotation in $SU(2)$
around a unit vector, $\mathbf{\hat{n}}$, by an angle of $2\phi$ may
be represented by
\begin{eqnarray}
R_{\hat{n}} (2 \phi) &=& e^{-i \phi \bm{\sigma} \cdot \mathbf{\hat{n}}} \\
	&=& \cos(\phi) I - i\sin(\phi) \bm{\sigma} \cdot \mathbf{\hat{n}}.
\end{eqnarray}
For a vector, $\mathbf{\hat{n}} = n_x \hat{\imath} + n_y \hat{\jmath}$
in the $XY$ plane, this rotation may be written
\begin{eqnarray}
R_{\hat{n}} (4 \phi) &=&
	\left[ \begin{array}{cc}
	\cos(2\phi) & -\sin(2\phi)(n_y + in_x)\\
	\sin(2\phi)(n_y - in_x) & \cos(2\phi)
	\end{array}\right]. \label{eqn:R}
\end{eqnarray}

Expressed as an operation in the pseudo-spin basis of $\ket{0_L} =
\ket{\gamma'}$ and $\ket{1_L} = U^{\dagger}\ket{\tau}$,
Eq. (\ref{eqn:groverIt}) yields
\begin{eqnarray}
Q' &=&	\left[ \begin{array}{cc}
	1-2|U_{\tau\gamma}|^{2} & 
	-2U_{\tau\gamma}^{*}\sqrt{1-|U_{\tau\gamma}|^{2}} \\
	 2U_{\tau\gamma}    \sqrt{1-|U_{\tau\gamma}|^{2}}
	& 1-2|U_{\tau\gamma}|^{2}
	 \end{array}\right] \\ 
   &=& \left[ \begin{array}{cc}
	\cos(2\theta) & 
	-\frac{U_{\tau\gamma}^{*}}{|U_{\tau\gamma}|} \sin(2\theta) \\
	 \frac{U_{\tau\gamma}}{|U_{\tau\gamma}|}    \sin(2\theta)
	& \cos(2\theta) \label{eqn:Q}
	\end{array}\right],
\end{eqnarray}
where we have defined the angle $\theta$ by
\begin{eqnarray}
\sin \theta &=& |U_{\tau\gamma}|, \label{eqn:theta1}\\
\cos \theta &=& \sqrt{1-|U_{\tau\gamma}|^2} . \label{eqn:theta2}
\end{eqnarray}
 
Comparing the equation for an arbitrary rotation, Eq. (\ref{eqn:R}),
and the equation for $Q'$, Eq. (\ref{eqn:Q}), we see that $Q'$ is a
rotation by an angle of $4 \theta$ around a unit vector defined by
\begin{eqnarray}
\mathbf{\hat{n}} &=& -\Im{\frac{U_{\tau\gamma}}{|U_{\tau\gamma}|}}
	  \hat\imath +\Re{\frac{U_{\tau\gamma}}{|U_{\tau\gamma}|}}
	  \hat\jmath \label{eqn:nhat}.
\end{eqnarray}
Clearly $\mathbf{\hat{n}}$ is a unit vector lying in the XY plane.

With appropiate definitions of $\theta$ [Eqs. (\ref{eqn:theta1}) and
(\ref{eqn:theta2})] and $\mathbf{\hat{n}}$ [Eq. (\ref{eqn:nhat})] we can
express $Q'$ as
\begin{equation}
Q' = e^{-i 2\theta \mathbf{\hat{n}} \cdot \bm{\sigma}}.
\end{equation}
This shows that $Q'$ is a rotation in SU(2). The angle rotated depends
only on the magnitude of a single matrix element,
$U_{\tau\gamma}$. The direction of rotation is only dependant on the
phase of the same matrix element, $U_{\tau\gamma}$. This equation holds
for arbitrary numbers of qubits, and for arbitrary choice of $U$.

\begin{figure}[ht!]
\begin{center}
\scalebox{0.6}{
\input{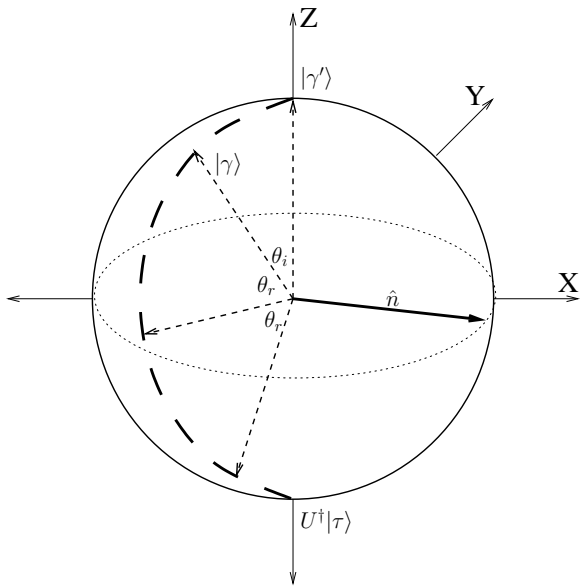}
}
\caption{Bloch sphere showing axis of rotation} \label{fig:Bloch}
\end{center}
\end{figure}

We now find the probability of success of Grover's algorithm. First,
we consider the initial state $\ket{\gamma}$. In terms of a rotation
we see from Eq. (\ref{eqn:gamma}) that
\begin{eqnarray}
\ket{\gamma} &=& \sqrt{1-|U_{\tau\gamma}|^{2}} \ket{\gamma'} +
U_{\tau\gamma} U^{\dagger} \ket{\tau} \\
	&=& e^{-i \theta \mathbf{\hat{n}} \cdot \bm{\sigma}} \ket{\gamma'}.
\end{eqnarray}
The matrix $e^{-i \theta \mathbf{\hat{n}}\cdot \bm{\sigma}}$
represents a rotation of $2 \theta$ around the $\mathbf{\hat{n}}$ axis,
Eq. (\ref{eqn:nhat}), on the pseudo-spin Bloch sphere. Initially we are
rotated by an angle of $\theta_i = 2 \theta$. Each application of $Q'$
rotates us by a further $\theta_r = 4 \theta$. Every rotation is
applied in the same plane, orthoganal to $\mathbf{\hat{n}}$. This is shown in
Figure \ref{fig:Bloch}.

After $s$ applications of the Grover iteration, the state of the
system is
\begin{eqnarray}
\ket{\psi_s} &=& e^{-(2s+1) \theta \mathbf{\hat{n}} \cdot \bm{\sigma}}
\ket{\gamma'} \\
	&=& \cos[(2s+1) \theta] \ket{\gamma'} \nonumber \\
	& & + \frac{U_{\tau\gamma}}{|U_{\tau\gamma}|} \sin[(2s+1) \theta]
U^{\dagger}\ket{\tau}.
\end{eqnarray}
The probability of success $P_s$ is simply given by the absolute value
squared of the amplitude of $U^{\dagger}\ket{\tau}$. We then find
\begin{equation}
P_s = \sin^2{\left[(2s + 1) \theta \right]}, \label{eqn:ps}
\end{equation}
where $s$ is the number of applications of the Grover iteration, and
$\sin \theta = |U_{\tau\gamma}|$. Specifically in the case under
consideration by Feng using $X$ rotations, we have $U=W_n$,
Eq. (\ref{eqn:W}). For an arbitrary prepared state $\ket{\gamma}$ and
arbitrary marked state $\ket{\tau}$, we obtain
\begin{equation}
\sin{\theta} = |\braopket{\tau}{W_n}{\gamma}| = \frac{1}{\sqrt{N}},
\label{eqn:theta}
\end{equation}
where $N = 2^{n}$ and $n$ is the number of qubits. The probability of
success given in Eq. (\ref{eqn:ps}) and Eq. (\ref{eqn:theta}) is the
same as may be obtained for Grover's algorithm based on the
Hadamard transformation.

The authors would like to thank Gerard Milburn for support. This work
was partly supported by the Australian Research Council, the
Australian government and by the US National Security Agency (NSA),
Advanced Research and Development Activity (ARDA) and the Army
Research Office (ARO) under contract number
DAAD19-01-1-0653. H.S.G. would like to acknowledge financial support
from Hewlett-Packard.

\bibliography{groverWrong} 

\begin{thebibliography}{4}
\expandafter\ifx\csname natexlab\endcsname\relax\def\natexlab#1{#1}\fi
\expandafter\ifx\csname bibnamefont\endcsname\relax
  \def\bibnamefont#1{#1}\fi
\expandafter\ifx\csname bibfnamefont\endcsname\relax
  \def\bibfnamefont#1{#1}\fi
\expandafter\ifx\csname citenamefont\endcsname\relax
  \def\citenamefont#1{#1}\fi
\expandafter\ifx\csname url\endcsname\relax
  \def\url#1{\texttt{#1}}\fi
\expandafter\ifx\csname urlprefix\endcsname\relax\def\urlprefix{URL }\fi
\providecommand{\bibinfo}[2]{#2}
\providecommand{\eprint}[2][]{\url{#2}}

\bibitem[{\citenamefont{Feng}(2001)}]{Fen01}
\bibinfo{author}{\bibfnamefont{M.}~\bibnamefont{Feng}}, \bibinfo{journal}{Phys.
  Rev. A} \textbf{\bibinfo{volume}{63}}, \bibinfo{pages}{052308}
  (\bibinfo{year}{2001}).

\bibitem[{\citenamefont{Berman et~al.}(2000)\citenamefont{Berman, Doolen, and
  Tsifrinovich}}]{BDT00}
\bibinfo{author}{\bibfnamefont{G.~P.} \bibnamefont{Berman}},
  \bibinfo{author}{\bibfnamefont{G.~D.} \bibnamefont{Doolen}},
  \bibnamefont{and} \bibinfo{author}{\bibfnamefont{V.~I.}
  \bibnamefont{Tsifrinovich}}, \bibinfo{journal}{Phys. Rev. Lett.}
  \textbf{\bibinfo{volume}{84}}, \bibinfo{pages}{1615} (\bibinfo{year}{2000}).

\bibitem[{\citenamefont{Sorensen and Molmer}(1999)}]{SM99}
\bibinfo{author}{\bibfnamefont{A.}~\bibnamefont{Sorensen}} \bibnamefont{and}
  \bibinfo{author}{\bibfnamefont{K.}~\bibnamefont{Molmer}},
  \bibinfo{journal}{Phys. Rev. Lett.} \textbf{\bibinfo{volume}{82}},
  \bibinfo{pages}{1971} (\bibinfo{year}{1999}).

\bibitem[{\citenamefont{Grover}(1998)}]{Gro98}
\bibinfo{author}{\bibfnamefont{L.~V.} \bibnamefont{Grover}},
  \bibinfo{journal}{Phys. Rev. Lett.} \textbf{\bibinfo{volume}{80}},
  \bibinfo{pages}{4329} (\bibinfo{year}{1998}).

\end{thebibliography}

\end{document}